# The Standard Problem

*Enrico Coiera*
*Centre for Health Informatics*
*Australian Institute of Health Innovation*
*Macquarie University*
*Sydney, Australia*
*enrico.coiera@mq.edu.au*

*Draft 16 August 2023*

**Abstract**:

**Objective**: This paper proposes a framework to support the scientific research of standards so that they can be better measured, evaluated, and designed.

**Methods**: Beginning with the notion of common models, the framework describes the general standard problem - the seeming impossibility of creating a singular, persistent and definitive standard which is not subject to change over time in an open system.

**Results**: The standard problem arises from uncertainty driven by variations in operating context, standard quality, differences in implementation, and drift over time. As a result, fitting work using conformance services is needed to repair these gaps between a standard and what is required for real-world use. To guide standards design and repair, a framework for measuring performance in context is suggested, based on signal detection theory and technomarkers. Based on the type of common model in operation, different conformance strategies are identified: (a) Universal conformance (all agents access the same standard); (b) Mediated conformance (an interoperability layer supports heterogeneous agents) and (c) Localized conformance (autonomous adaptive agents manage their own needs). Conformance methods include incremental design, modular design, adaptors, and creating interactive and adaptive agents.

**Discussion**: Machine learning should have a major role in adaptive fitting. Research to guide the choice and design of conformance services may focus on the stability and homogeneity of shared tasks, and whether common models are shared ahead of time or adjusted at task time.

**Conclusion**: This analysis conceptually decouples interoperability and standardization. While standards facilitate interoperability, interoperability is achievable without standardization.

**Introduction**

Standards are a foundational preoccupation in the technological realm. Adopting a standard can help meet performance benchmarks for a technology, such as its safety, efficacy, or reliability. Standards also allow us to craft a technological artefact so that it will fit, or interoperate, with others, and opens technology creation to competition and independence from any one manufacturer.

Standards thus guide the construction and operation of machines, software, and human processes. Globally, most standards activity focuses on the complex and challenging task of forging agreement between different parties on whether a standard is needed, what it should define, and how it should be communicated, implemented and enforced. The business of standardization is thus one of consensus, design, engineering, regulation, politics, pragmatics and lobbying.

The academic literature on standards has a long history, is large, and is scattered across disciplines from economics[1] to information system design[2]. In health informatics, most focus has been on the specific issue of semantic interoperability to facilitate machine to machine communication[3][4]. Such standards may support low level data interchange between information systems, as well as ensure they recognise and process concepts and their relationships in similar ways[5].

Crafting, adhering to, and maintaining standards remains an ongoing and unsolved challenge – the standard problem. Over their lifetime, standards seem to need repair or augmentation. There can be dissatisfaction when some feel a standard is illogical or constrains activity too much, whilst others feel it does not go far enough[6][7]. When different organisations implement an identical standard, there are often differences in how that is done, hampering interoperation. Over time, new standards can emerge to compete with old standards.

**Crafting a research agenda for the standard problem**

The general standard problem is the seeming impossibility of creating a singular, definitive and static standard in real-world settings. There are three broad responses to this problem. The first sees standards as inherently perfectible, and any need to alter a standard as an act moving closer towards that perfection. A second sees the need to make accommodations to the design of a standard as a pragmatic response to "real world" problems external to the standard, such as vagaries of the marketplace or the rapid rate of change in technology[8]. A third perspective sees these challenges as manifestations of the underlying nature of standards and thus inevitable in the process of standardization.

Indeed, standards must be a researchable construct. Standards must have common properties, and those properties must influence real world outcomes in explainable ways. Yet the scientific study of standards and standardization has attracted much less attention than pragmatic global efforts of standard creation, adoption, and interoperation. However not everything can be meaningfully standardized. Not everything necessarily should be. We do need formal approaches to answer core questions about when (and when not) to standardize, what to standardize, and how much of any standard one should comply with.

This paper proposes a programmatic approach to standards research. The general standard problem is decomposed into several inter-related research subproblems:

- **What is a standard?** To study standards, we require a conceptual framing that allows development of theory and experimental testing. Building on prior work[9-13], this paper lays out a theoretical framing based on common models shared between agents, with a focus on information standards.
- **Why do standards vary?** The general standard problem manifests as variation in standards over time and space (different locations may vary in how standards are implemented). While standards are a technology, they are embedded in the real world, and any account of the causes of variation must adopt a socio-technical lens.
- **How is standard performance measured?** Researching standards requires metrics that permit quantifiable comparison. The socio-technical nature of performance means measurement must account for variation in different contexts.
- **How can standards be adjusted to manage the impact of unwanted variation**? As contexts, tasks and entities change, changes need to be made to standards or their implementations. Research should help identify different "repair" options.
- **What is the best approach to work with multiple standards?** When multiple standards co-exist, there is a need to translate between them for tasks to be solved co-operatively. Hand-crafting such mappings is not a scalable solution in complex settings. Research can identify different solutions to mapping between standards and the best strategies for maintaining interoperability.
- **Are there alternate ways to conceptualise the problem that standards seek to solve?** Requiring technologies to adhere to a standard has proven an effective approach to building technology ecosystems but has not made variation in technology design go away. There thus remain foundational questions about the standards enterprise. For example, the relationship between interoperability and standardisation is worthy of reappraisal. If we separated these two concepts, would that yield different approaches to technology design and use?

**What is a standard?**

To arrive at a conceptual framing of standards that permits theory development and experimental testing, we start with a first principles description of the purpose of standards. Many tasks require the co-operation of entities to achieve goals that cannot be achieved individually. These entities might be physical, such as parts of a machine or molecules in a metabolic process. For information technology, these entities may be different elements of software or hardware. Sometimes they will have some independent agency, such as human or software agents. Co-operation requires that each entity is capable of successfully interacting with others to achieve a common goal. *Interoperability* is the degree to which interacting entities can successfully complete a co-operative task[14]. By corollary, *disinteroperabilty* would be the degree of difficulty experienced when trying to interact.

For interoperation, entities must somehow share a *common model* of how to interact with each other, sufficient to accomplish the shared task. A common model might be expressed through an entity's structure, composition, or function. For example, the connecting parts of two machines must physically model a common boundary shape to fit each other. Humans must share language and knowledge of the world, or common ground, to co-ordinate themselves[10]. Computers must represent data input and output in a manner recognisable by other computers.

*A standard is a common model that is agreed ahead of time*. It describes a process, structure or function that must be conformed with, for a specified coordinative task to be completed successfully. A standard is thus by definition a predetermined restriction or constraint on an entity's allowable boundary forms, internal composition or behaviour, with the expectation that the cost of reduced choice brings co-operative benefit. As we will see later, this formulation allows for common models which are not standards, and by inference, interoperation without standards.

With a fully described and deterministic closed system, with no external interactions, we can have perfect knowledge and define exactly the standard needed to produce the best outcome. For example, in a deterministic computational simulation, rerunning the same experiment with the same initial conditions should always arrive at the same outcome. In reality, the universe is incompletely known and described. Standards do not exist in isolation but are part of a larger, imprecisely understood, socio-technical system[15]. The unpredictable nature of large complex and dynamic systems means there is inherent uncertainty about the nature and purpose of tasks, the entities that participate in a task, and their cooperative interactions.

This leads to the *general standard problem conjecture* – in an open and dynamic complex system (one which is porous and interacts with the world beyond it), it is not possible to create a singular, persistent and definitive standard which is not subject to change over time. This conjecture could be disproved by identifying standards which, over a reasonable period, do not require updating, do not exhibit variant forms, and do not have independently competing standards. Determining a 'reasonable period' for the test is critical. For example, the bayonet light bulb socket was patented circa 1870, and it took about 12 years before Edison patented his competing screw socket. After 150 years there are now hundreds of socket types[16]. So, to test the conjecture with a new standard, one could estimate the expected time until variation should appear, based upon comparable pre-existing standards. If persistently invariant standards do exist, they would be the object of intense study. Do they exist only because their domain of use is highly stable, or because something in their design or manner of use is so different to standards which do exhibit variation?

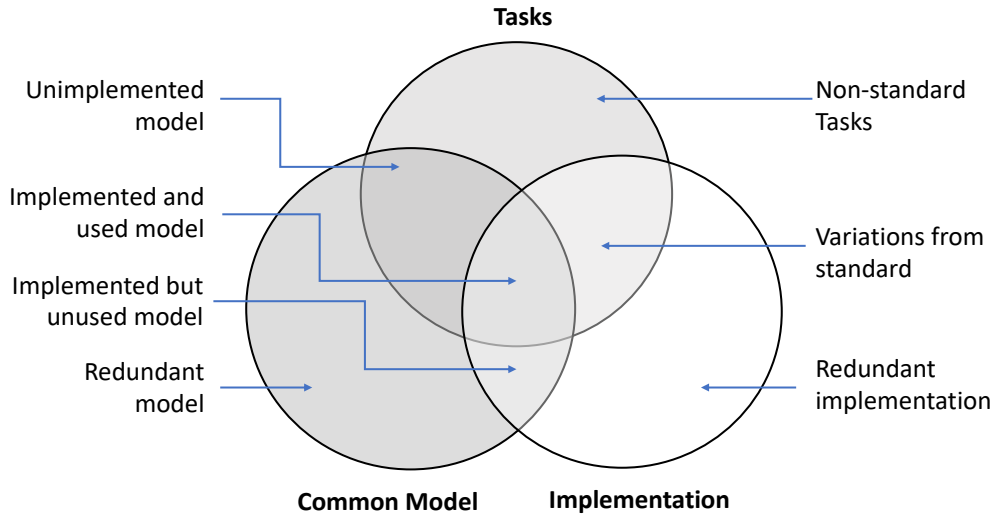

*Figure 1*: The standard problem is manifest as differences between what is standardized in the common model, how much of that model is implemented in the real world, and the purposes to which an implemented common model is applied.

**Why do standards vary?**

Core to the general standard problem conjecture is the idea that we will see variations in standards over time and space, often manifest as differences between a standard, its real-world implementation, or how tasks are undertaken (Figure 1). Studying standards as they operate in a socio-technical world requires analysis of the origins of variation, which clearly has many causes, is dynamic and evolving:

- *Variation in operating context:* The world for which a standard is designed will never be exactly like the world within which it is used. Variations in physical and digital systems, priorities, resources, people, processes and tasks exist even between very similar environments. New technologies with new functionalities beyond the scope of a standard can appear more quickly than a deliberative standards development process. Designs that once worked well may exhibit diminishing returns because of such accumulating changes.

- *Variation in standard definitions*. There is foundational uncertainty about how to describe physical or informational objects. Even obvious concepts such as 'chair' are fluid and matching a concept label to real world examples is probabilistic[17]. This semantic heterogeneity means that information standards defining similar concepts will likely differ because there is no semantic 'ground truth'. For example, there are multiple overlapping standards for the structure and communication of medical data, and multiple versions of individual standards can be in use at the same time. The history of standards is full of this type of variation, from differences in rail gauges through to videotape formats[18]. Perceived deficiencies in one standard trigger development of an alternate one by competing groups.

- *Variation in standard quality*: Like any artefact, standards may contain errors and omissions[19], and the risk of errors is likely to increase with the complexity and size of a standard. The adoption of high-quality processes during standard development and refinement should minimize the rate of such errors. Automated mechanisms to identify errors in standards can also assist[20][21].

- *Variation in standard implementation*: There is uncertainty about how any entity has embodied a standard. Standard takers might choose to only conform to a subset of a standard, reflecting expectations about which parts are most relevant. Implementation differences can arise when a standard is defined ambiguously and permits some freedom in interpretation, or through errors in interpretation. Consequently, variation in the way a standard is implemented is the norm. For example, the implementation of information and document standards by different organisations often contain errors and variations in expression[22].

- *Standard drift*: All these variations unfold dynamically and accrete. Indeed, all natural systems accrue local variation and increase in complexity over time[23]. As a result, the 'fit' between a standard and the world it was intended for drift apart with time, as will the effectiveness of any co-operative task that depends on a standard. Standard drift may manifest at the edge of a standard. Information standards for example create residual categories such as ''not elsewhere categorized'', ''none of the above'' or ''not otherwise specified'[24]. With time, the phenomena assigned to these categories may spawn new conceptual classes, pushing a standard's boundary further out.

**How should a standard's performance be measured?**

Given that variation in standards is expected, so too should variation in standards performance. Researching standards thus requires metrics that permit quantifiable comparisons of performance. The socio-technical nature of standards means that it is not possible to discuss the performance of a standard without also discussing the way it has been implemented and the context within which it works. For example, conformance testing of an implemented standard should not just be a bench test against pre-determined cases but should reflect the actual distribution of real-world tasks. Signal detection theory provides one general approach to measurement that is sensitive to context. In the diagnostic setting, biomarkers are used to decide if a disease is or is not present in an individual. Box 1 contains a proposal to utilise *technomarkers,* which by analogy provide measures of whether a standard is effective or not in a context. The choice of the best technomarkers for a given standard, context or task, and how technomarkers (individually or as a panel) are best measured and analysed, are open research questions.

**How can standards be adjusted to manage the impact of unwanted variation**?

Given the seeming inevitability of standard variation, there likely will always be a *practice-standard gap*, where a standard cannot fully satisfy the purposes for which it was created. Similarly, there will be an *interoperability gap* between different standards. *Fitting work* refers to the efforts needed to bring the common models of interoperating entities into closer alignment and may involve adjusting a standard, its implementation, or the target task.

Indeed, performance adjustments are characteristic of all work, not just standards, and are seen as unavoidable, ubiquitous and necessary[11 25]. The process of implementation itself has long been seen as a process of mutual adaptation of technology and organization[26]. Fitting work is therefore a process of unfolding or emergent design that occurs post-implementation[27 28].

While it is common to talk about a one-way flow of information from *standard makers* (who define a standard) to *standard takers* (who implement and use a standard), information may flow in both directions (Figure 5). Those who define a standard might receive information from standard users who have identified deficiencies in it, triggering a revision of the standard. It is the direction of information flow that determines who is a 'taker' or 'maker'.

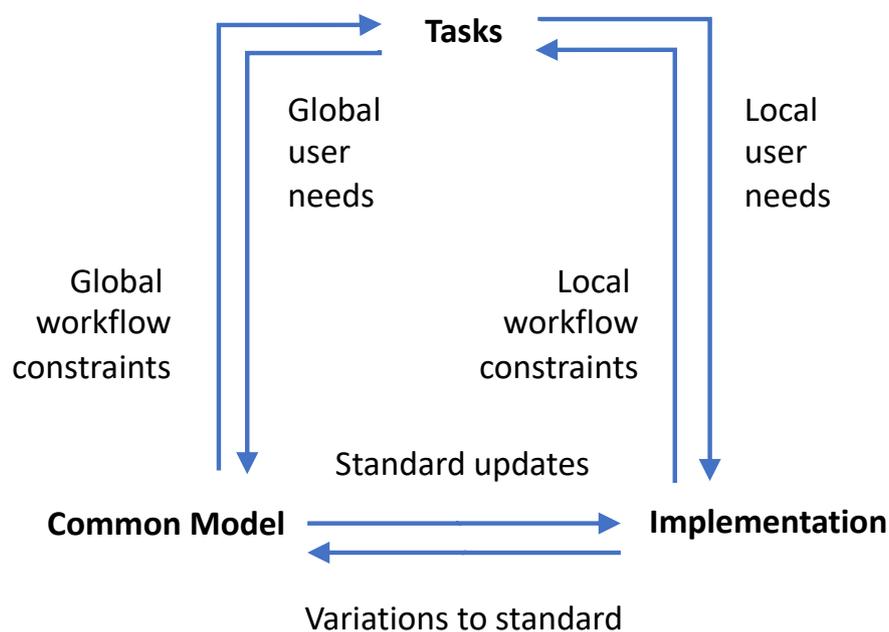

*Figure 5*: *Fitting work seeks to better approximate common models with their implementations and target tasks. Information about practice and interoperability gaps may lead to changes in standards, their implementations, or the tasks themselves.*

Understanding where fitting work is needed, how it is done, and whether it has succeeded, are critical to understanding how standards evolve, and how change-management is best accomplished. Fitting work may modify a standard, its implementation, or target tasks:

- *Task fitting*: Practice gaps document the differences between "work as done and work as imagined" in a standard. Humans will work around standardized processes by adding, changing, or removing steps and such changes signal altered task requirements. Workarounds are thus local 'repairs' to standardized processes that can provide missing information on local needs back to standard designers[11]. For example, human users may deviate from the standardized recommendation of computerized decision support system to suit novel patient-specific circumstances. The reverse is also possible, with local tasks being adjusted to better accommodate a standard. For example, a human workflow for hospital medication administration may be reshaped to conform to the workflow of a new electronic medication management system.

- *Implementation fitting*: Intentional deviations from a reference standard in a local implementation may signal local needs or constraints not envisaged within a standard. For example, a list of codes used to categorize patients might be augmented to reflect specific local information systems, regulations or practices.
- *Standard fitting*: Changes to a standard's definition (e.g. by an international standards body) are driven by aggregate assessments of the changing needs of implementors or users, or by anticipating new needs. For example, local implementation variations provide information to standard designers about the suitability and performance of their standard.
- *Interoperation fitting:* To engineer co-operation between two implementations of a standard, or two different but overlapping standards, fitting work ensures there is a common model for the target tasks. In general, any two standards may have a common core that is already shared, as well as specialized elements where their approach to task execution is unique (Figure 6(a)). Fitting work between standards need only occur for those elements necessary to complete a specified task (Figure 6(c)), rather than completely aligning the standards.

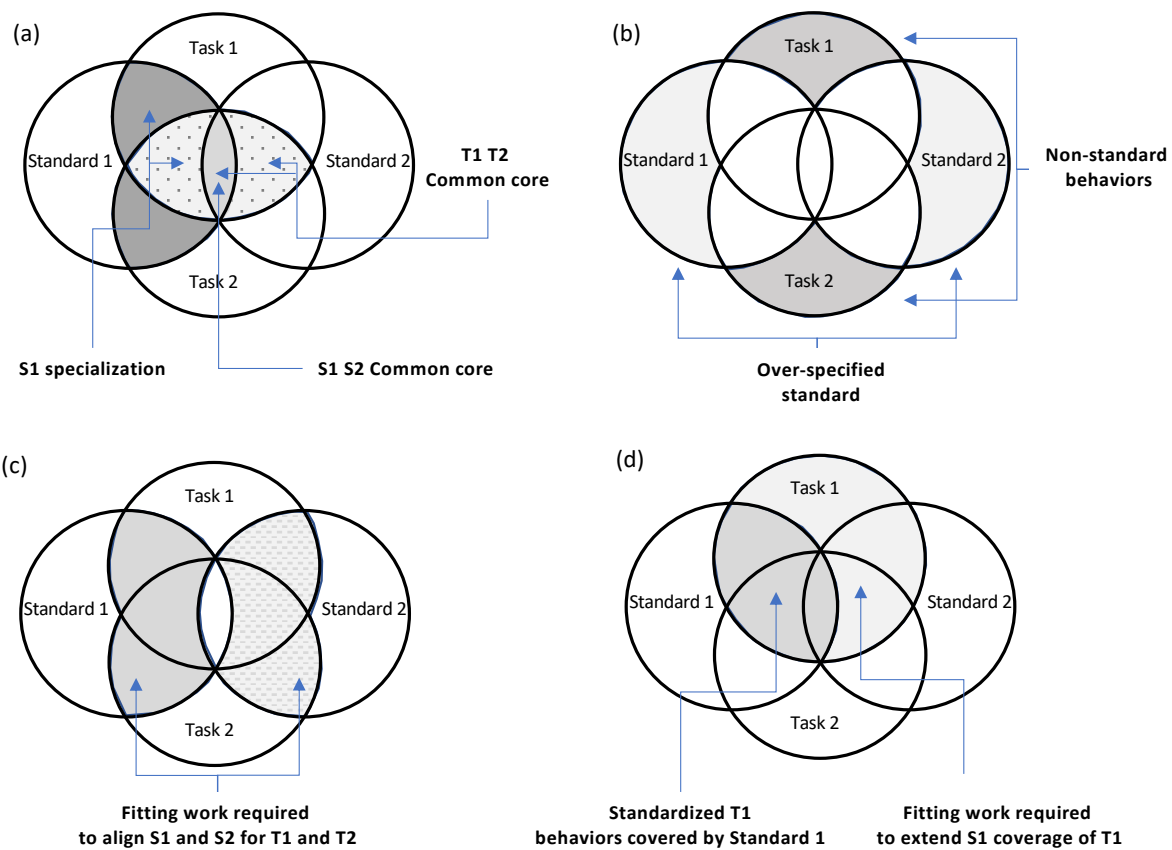

*Figure 6*: For two overlapping implemented standards S1 and S2 (e.g. two implementations of the same standard or two different standards) and two similar but not identical tasks T1 and T2 (e.g. the same task in two different settings or related tasks in the same setting) we can identify: (a) a standard common core which is identical cross S1 and S2, and a task common core where S1 and S2 effectively interoperate. For some portion of T1 or T2, only one of the standards provides coverage, hindering between standard interoperation (e.g. a *specialization in one of the standards); (b) Some elements of T1 and T2 will not be covered by any standard and some parts of S1 and S2 may not play a role in task support and interoperation for these tasks. If interoperation is desired between two standards, then where gaps exist either (c) fitting work is needed to align S1 and S2 or (d) to align a standard to any unserved parts of the task space.*

**What is the best approach to work with multiple standards?**

When interacting agents have adopted different standards for a shared task, interoperability is impeded, and is there a need to translate between them to support co-operation. This translation is not static, given the ongoing appearance of variations. The best approach to building and maintaining interoperability will likely depend on the nature of the cooperative task, the capabilities of the entities carrying it out and their sociotechnical context. *Conformance services* are the collection of methods, processes, and technologies used to ensure common models remain aligned.

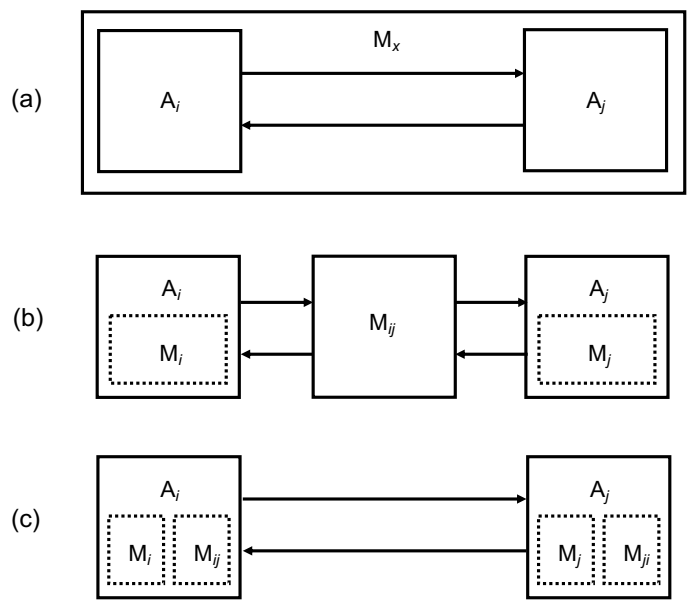

*Figure 7*: *Three models of conformance service: (a) Universal conformance: All agents have access to the same global standard ($M_x$) of interaction; (b) Mediated conformance: Adaptors provide an externally situated conformance service to interoperating agents; (c) Localized conformance: Autonomous adaptive agents internalize their conformance functions. Standards are mandated in (a), incompletely adhered to in (b) and potentially helpful but not necessary in (c).*

Three broad conformance service models appear possible, based upon the location of conformance services and common models, each demanding increased capability of the interacting entities (Figure 7):
- *Universal conformance:* In a fully deterministic setting, there need only be one standard, and the entities that use the standard could enact it perfectly. For example, multiple software programs in a single computational environment might all make different contributions to a simulation but share the same common model of how to communicate. When fitness gaps emerge, for example because the simulation must evolve to address a changing task set, updates to the common model are immediately available to all programs within the shared computational environment. This universal model is also well suited to open or non-deterministic "real world" settings when tasks are relatively stable over long periods. In such cases, testing can help determine if gaps emerge between a local implementation and a definitional reference standard.

- *Mediated conformance*: In open and non-deterministic settings, interacting entities do not always have identical common models, and may use an external conformance service to align their models. For information systems, these services might reside in middleware or an interoperability layer that sits between an operating system and various applications[5]. Within such a layer, *adaptors* act as mediators, and may facilitate both syntactic and semantic interoperability. For example, in a process known as mapping, an adaptor may translate both the format and labels produced by one information system into a format and labels that can be understood by another (for example utilising a reference terminology)[29].
- *Localized conformance*: Sufficiently sophisticated entities may have the capability to undertake conformance tasks themselves[30][31]. They may either explain their structure or behaviour to other agents or adapt their own behaviour to conform with others. A software agent might learn the best way to collaborate with another through a series of interactions, discovering a common communication protocol and identifying shared and unshared capabilities. This ongoing dialogue between entities seeks to create or refine their common ground, or common model[10].

Interoperability can be maintained by using one or a combination of these conformance models (e.g. standard plus mediation service, standard plus localized conformance).

Conformance services may exploit different technical approaches to attempt common model alignment. Well-crafted standards will exploit good design principles such as modularity in design to minimize fitting work, and more advanced services may use adaptive methods such as machine learning.

*Incremental Design*

The traditional approach to modifying an information standard is for a Standard Development Organization (SDO) to release revisions as it sees fit[5]. An SDO may receive feedback or change requests from standard implementors or users and decide which will be adopted. Standard updates should be guided by defined processes and rules. For example, the FHIR health information standard uses an 80/20 rule i.e. once 80% of standard implementations include the same non-standard element, then that element becomes formally incorporated into the standard[32]. By focussing only on current implementations, this ensures FHIR does not include rare boundary cases or ambit claims for novel elements that may be contested. It also empowers the deciding committee in the SDO to reject requests from non-representative but powerful organisations. When an SDO is captured by a particular group, change requests may not be processed in an unbiased manner. The change process has the capacity to be automated in part, and machine learning methods have been applied to identifying common system requirements from change request documents[33].

*Modular design*

Separating a technology design into discrete modules is a classic engineering approach to complexity reduction, sequestering different functions into separate modules ("separation of concerns"), where possible limiting the impact of changes to specific modules[34]. Making changes to the underlying architecture that defines how modules are to be built or connected can however have far reaching consequences. Modules can be created on the basis of conceptual independence of functions, or by tying to separate structure from function. Kernel design methods separate modules into core and non-core functions (e.g. sequestering functions with a short half-life for change). Standards such as ontologies and terminologies may use a semantic stack, where objects are grouped by semantic level or category e.g. atomic terms, concepts, rules and metarules, with each layer adding semantic complexity[35]. For example, terminologies may use combining rules to generate complex terms from a library of basic concepts (post-coordinated) instead of specifying all such terms explicitly (pre-coordinated).

*Adaptors*

An adaptor is any intermediary technology that connects otherwise dissimilar entities to facilitate a coordinative task. An adaptor contains a mapping mechanism to translate between different standards. Adaptor users do not need a common model of the adaptor, but the adaptor clearly must understand the entities it connects. Adaptors are thus liminal constructs that behave as *boundary objects*, sitting in the middle of intersecting worlds, attempting to satisfy the information requirements of each[36][37].

With information technology, adaptors may use explicit mapping rules and methods to translate data stored using one standard into a form understandable by a different standard[38][39]. Mapping across different terminologies is a routine task for developers of clinical information systems[40][41]. ETL (Extract, Transform, Load) methods are standard tools for ingesting data of different provenances into a repository such as a data warehouse[42], for example converting clinical databases into the OMOP standard[43]. Semantic heterogeneity across standards may be addressed by ontology matching, either using an alignment ontology or a suite of algorithms and heuristics to infer relations between concepts[44][45] e.g. "cardiac vasculature" might map to "vasculature of heart" and "fracture of left tibia" to the more general 'fracture of tibia'. As standards can come from very different world views, a direct one to one mapping is not always possible, in the same way that it is not always possible to fully translate the meaning of a phrase in one language into another.

While adaptors can solve the standard problem for the services that use them, they also push the problem up to the adaptor. Fitting work still needs to happen whenever mapping methods require updating. Modularity in the design of mapping methods may help minimize the cost of such changes.

A more flexible approach to adaptor design is to create the mapping service using machine learning, both for the initial map, and then to dynamically update it as the context changes. For example, natural language processing has relied on the use of probabilistic mappings between words (for example using word embeddings) rather than using standardized definitions[46]. Machine learning can also be used to discover mappings for ontology matching[47][48], including through the use of generative pre-trained transformers (GPTs)[49].

Machine learning thus permits the creation of emergent middleware[50], taking a dynamic approach to interoperability where some of the mediating infrastructure is generated as it is needed and specifically for the current context[51-53]. Unlike manually created mappings, machine learning has greater potential to scale, update dynamically, and potentially adapt to new domains.

*Adaptive agents*

Co-operating entities may be capable of undertaking fitting work on their own, without recourse to a mediating conformance service. For example, software agents may possess discovery capabilities similar to those found in a mediating interoperability layer. This may be necessary when agents are not implemented on the same middleware, or the environment is sufficiently dynamic that reliance on a mediator is not feasible.

Whilst adaptors implement a one-way process of discovery (the adaptor tries to understand how co-operating entities function), independent agents can participate in a two-way dialogue. For example, two agents may 'speak' different clinical languages (use different terminological sets) but can interact with each other to agree upon the mapping they will use to associate their dissimilar terms. They might do this by labelling a set of shared examples ("I call this an adenocarcinoma"). Agents can test the common language, knowledge and assumptions between them, and then attempt to modify deficiencies in their shared understanding to ensure their interactions succeed.

Multi-agent systems may employ a variety of methods to reach consensus, for example through negotiation and argumentation protocols [54-56]. Such grounding is not 'frictionless' however, because the environment external to agents imposes constraints and costs. For example, software agents may need to access external services or accommodate limits on computational or communication resources[57][58]. The cost of grounding (or grounding efficiency) is also shaped by the degree to which agents share a similar common model (homophily)[13]. The more similar the agents, the easier it is for them to engage, and the more "personalized" is the explanation.

Model discovery by agents is a focus of study in several other fields, and these may assist in designing adaptive conformance services. In cryptographic model extraction games, a software adversary tries to discover the model within another entity. By using a Model Extraction Attack, the aim is to efficiently and effectively reconstruct a model close to the entities actual functionality, based on a sequence of query-response pairs [59]. An adversary can for example exploit deep learning on labels obtained from a classifier under attack, and build a functionally equivalent classifier without knowing the type, structure or underlying parameters of the original classifier, with high accuracy[60]. Making artificial intelligence explainable can also be framed as model discovery by agents[61][62]. Explanations will sometimes be necessary for building shared understanding e.g. Asking "Explain why you classified this condition as 'IDDM' and not 'Type II Diabetes'?" might reveal deeper semantic information about an agent's knowledge. More recent developments with GPTs, not just for text but images and other data types, and enable single or zero shot domain adaptation[63][64], and will likely be transformational in adaptive agent design and interoperation.

**How can we select the right conformance strategy?**

Choosing between conformance service strategies requires an assessment of the stability and uniformity of the domain being modelled (Table 1). There is a choice between how much of the common model is agreed ahead of time (e.g. in the form of a standard), and how much is left to be developed at the time it is needed:

- Relying on a single universal standard makes sense in relatively stable and homogenous contexts where interoperability gaps are likely to be small and to arise infrequently, minimising fitting work. Additionally, the scope of shared tasks needs to be bounded, feasible and their standardization cost-effective. Choosing universal standardization should also become more compelling as the number of entities required to interact increases. Monitoring standard performance over time will be needed to allow detection of new edge cases and performance drift, triggering updating of the universal standard.
- Adaptors make sense when multiple standards co-exist, the interoperability gap between them is substantial, but the task domain is relatively stable and homogenous, making the time and effort devoted to crafting adaptors worthwhile. Within the field of distributed software systems, for example, it has long been recognised that any approach that assumes a common standard is destined to fail[44].
- Localized conformance through the dynamic adjustment of common models shared by interacting entities is needed when shared tasks are uncertain, highly diverse, unique, or standards are unavailable[10]. Local dynamic adjustments are also worthwhile when their cost is small compared to the impost of standardization.

Choosing between conformance service strategies thus requires a cost-benefit trade-off analysis to identify the minimum amount of effort needed to create and maintain a standard versus the cost of undertaking fitting work beyond the standard (Figure 8). The shape of the cost curves for standardization and fitting work will vary with task, context and implementation, and the design of a standard. The ideal balance of strategies minimizes the total cost of standardization and non-standardized fitting work. When entities collaborate on fitting work locally, each may have a different cost curve, and could negotiate which undertakes the work to fit to the other. The *law of the mediated centre* states that interactions between entities should be driven to an intermediate level of model sharing over a series of interactions, where the total cost of maintaining shared models and constructing models at interaction time is minimized[13]. This mediated centre represents an interaction equilibrium point.

|  | **Stable world** | **Intermediate stability** | **Dynamic world** |
|---|---|---|---|
| **Homogenous world** | Universal standard | | |
| **Intermediate variability** | | Mediation services | |
| **Heterogenous world** | | | Localized conformance |

*Table 1*: The choice of conformance service is dependent on the stability and homogeneity of the shared tasks described in a standard, as well as the degree to which modelling these tasks is bounded, feasible and cost-effective. The rate of change in world, and between and within standards created for that world, shapes whether we standardize everything ahead of time or address gaps in near real time.

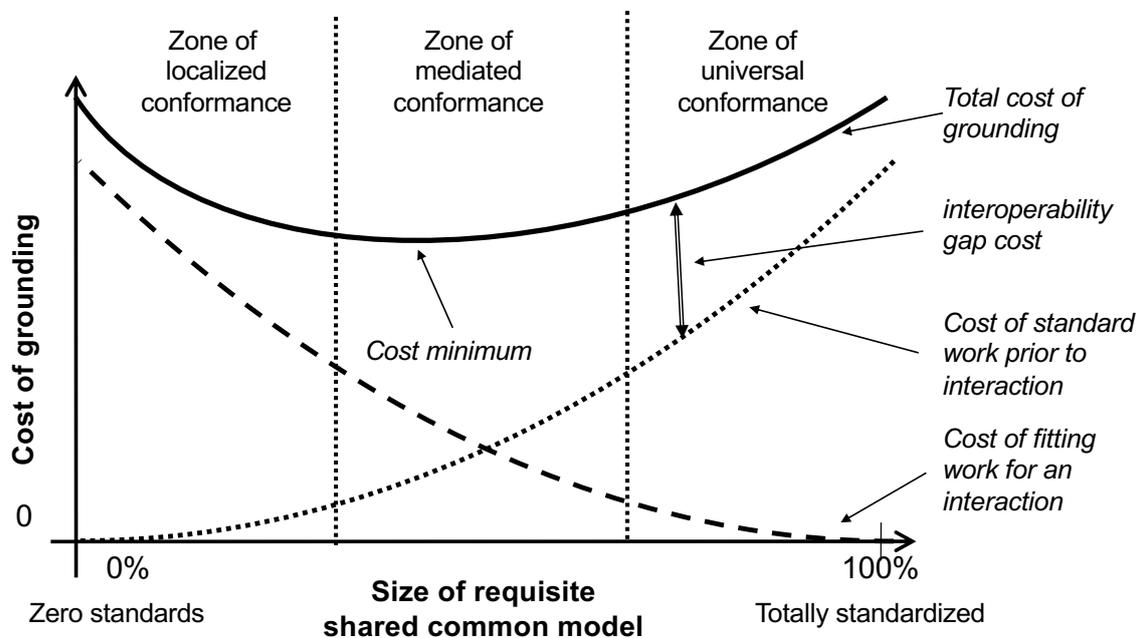

*Figure 8*: The choice of conformance service is dependent on the balance between the degree to which common models are shared ahead of time in a standard or need to be adjusted at task time. The ideal balance of conformance services minimizes the total cost of standardization and non-standardized fitting work (Adapted from Coiera, 2001)[13].

Artificial intelligence techniques such as GPTs offer new approaches to creating flexible and dynamic adaptors or autonomous agents capable of grounding. We can also contemplate using the same approach to create flexible and dynamic standards. When standards are adaptive, the adaptor becomes the standard. For example, a standard may allow interacting entities to initially only be loosely coupled, and the prescribe how at task time they should adapt to each other, making fitting work part of the standard. At the extreme, a fully adaptive standard is no longer a standard, as it asks nothing to be pre-agreed.

Finally, a resilient standard would exhibit graceful degradation as the contexts in which is used become more heterogenous over time. We would expect approaches to conformance that emphasize flexible and dynamic fitting work to exhibit greater resilience than static standards or conformance services. Research into fitting strategies would benefit from understanding the impact of adaptation in other complex systems such as biological ecosystems. Evolutionary adaptation is richly explored and can help explain when adaptations bring additional fitness to an organism or ecosystem, and when ongoing adaption can lead to reduced fitness[65].

**Are there alternate ways to conceptualise the problem that standards seek to solve?**

Standards are such a central construct in the technological enterprise that it is hard to envision other ways to arrive at interoperation between co-operating agents. However, in the real world, it seems no standard is immortal or unchanging. To paraphrase an old adage – all standards are probably wrong, none are perfect, but at any particular point, some are more useful than others. Given the apparent inevitability of the standard problem and the limits it imposes on successful interoperation, it is reasonable for research to try to reconceptualise the foundational problem of interoperation. That reconceptualization might begin by asking "Do we need standards to interoperate?".

Typically, standardization is considered essential for interoperability and this has been a central dogma of information system design[14]. Yet in open worlds, the common model between any two entities will likely include standardized and non-standardized components. When faced with highly heterogenous and dynamic environments, co-operating entities may need to find ways to interoperate with little standardization available to support them. In the complete absence of standards, interacting entities would be free to interoperate in any manner they wish.

We thus can conceptually *decouple the concepts of interoperability and standardization*. Not all common models are standards, but all standards are common models. While standardization facilitates interoperability, interoperability may be achieved without standardization[66]. This perspective allows us to move from focusing solely on standards as pre-agreed concepts to seeing interoperation as a dynamic, and potentially task time process.

What would things look like in a zero standards world? Firstly, from the perspective of an autonomous and adaptive entity, we would see standards for what they are - a workaround when entities cannot adapt. Standards are a technology of convenience, not a law of nature. To function without standards, there still must be some sharing of abilities to communicate, construct, learn and maintain common models. The hypothesis of linguistic innateness, often ascribed to Chomsky[67], states that humans have an innate capacity to recognize, learn and use language. Similarly, the visual system is hardwired to detect and learn from specific features[68]. Machine learning is also highly shaped by assumptions about what features are, how they might relate, and what the learning goal is[69].

**Conclusion**

Just because standards are everywhere and commonplace does not mean we understand them well, or that we cannot imagine a world in which we think about interoperability in a different, more adaptive way. The standard problem is an inevitable consequence of working within complex systems, and tackling it directly, rather than working around it, may allow us to reimagine how technology ecosystems work. The standardization process has served us well, but its limitations have become clear as the world becomes more complex. A robust, replicable, and scientific exploration of interoperability, both with and without standardization, may prove transformational.

Box 1: **How is standard performance measured?**

It is one thing to measure how faithfully a specific standard's implementation replicates a gold standard (e.g. through conformance testing). It is another to measure whether local variations from a gold standard are harmful or beneficial. To measure how well an implemented standard fits the needs of a specific context requires testing to move from assessing conformance to measuring performance. Performance testing requires evaluation of a standard to include the system of people and technologies of which it is but one component.

*Measuring Task-Standard Fit*

Signal detection theory (SDT) measures how well a test can distinguish two populations (e.g. with a disease and without a disease), and also shows how that test can be calibrated to best distinguish the populations[70]. We can borrow and generalize from the mechanics of SDT to guide the application of standards in a context. Specifically, we can measure how well we can distinguish between two populations of tasks (suitable and not suitable for a given standardised method). We can also identify how best to apply (or calibrate) the local use of a standard to maximise its appropriate application. For example, from a population of reports generated by an information system, STD should tell us how well one can distinguish which reports will be successfully encoded by a messaging standard for transmission, and which will not.

To assess the fit of a specific standard to a local context, we begin by obtaining a population of similar tasks which contain sufficient variation within them that not all can be completed in a standardised way. Next, unless the standard is applied indiscriminately, we need to identify a decision criterion to know when to apply a standard to a task in this population. In the SDT diagnostic setting, one typically chooses the value of a *biomarker* to decide if a disease is present or not (such as the blood concentration of a molecule). In the standards setting, we can proceed in a similar fashion, by selecting a *technomarker* whose value is used to decide the suitability of a task to be undertaken in a standardised way, in a specific context.

A technomarker, like a biomarker, should be a continuous variable that is at least strongly associated with successful task completion, and ideally would be causal and independent. For example:
- Clinical staff may employ a standardized workflow to administer medications, but occasionally deviate from standard using a workaround. This may be because many infrequently administered medications do not easily fit the standard workflow. To distinguish between common and infrequent medicines, one could use the historic rate of administration of medications at a health service to distinguish which medication tasks are suitable for the standardised workflow.
- A syntactic standard that encodes data (such as a messaging standard) may fail to correctly encode some variations in data values or sequence that occasionally appear in clinical documents. One could use an estimate of the level of noise in a document as a technomarker to decide whether to encode a document or not. Low levels of noise should see high performance by the standard but high noise levels may indicate that the standard is inapplicable. Documents with a greater than a certain level of noise would be treated differently to avoid the risk of error.
- A semantic standard such as an ontology may be used to identify the concepts within a clinical document. The percent of tokens in a document convertible to a concept could be used as the variable that distinguish which documents within a population are suitable for semantic analysis. Too few token conversions and the standard may incorrectly interpret a document's concept space, while many conversions should indicate that the standard is performing well.

To obtain our measures of standard-task fit, we create a gold-standard task population where tasks are labelled as suitable or unsuitable for application of a given standard. Next, we select a specific value of the chosen technomarker (the cut off value) above which we apply the standard (Figure 2). Counts are then made of those tasks to which the standard is applied (positive cases), and those where it has not (negative cases). Counts for true and false positives (the standard is applied, correctly or incorrectly) and true or false negatives (the standard is not applied, correctly or incorrectly) yield measures of a standard's precision and recall at a cut-off value. Different cut-offs will yield different performance. Too low a cut-off may yield too many errors as the standard is applied too widely (over-general). Setting the cut-off value to 0 is equivalent to ignoring variation and applying a standard universally (Figure 3 (a)). Too high a threshold may be unnecessarily conservative (over-specific) and minimize the benefits of the standard (Figure 3 (b)). Poor choice of a technomarker will make it hard to know when to apply the standard (Figure 3 (c), (d)).

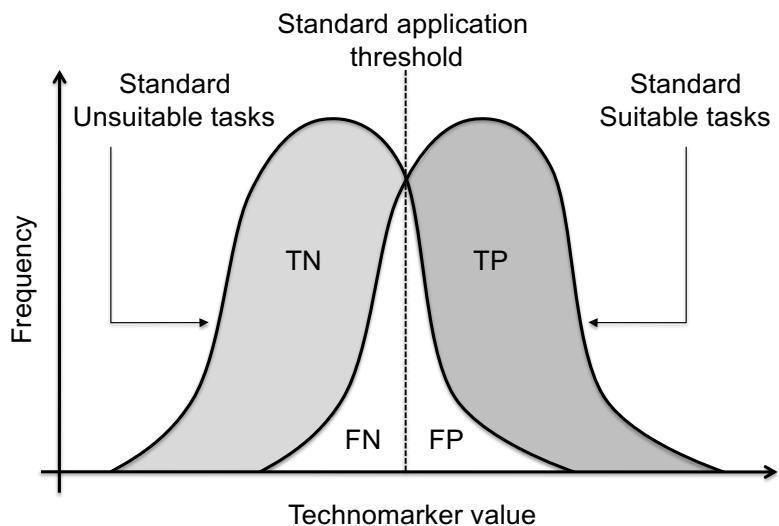

*Figure 2: Measuring how well a standard can be applied to real world tasks yields values for true positives (TP: the standard is appropriately applied), true negative (TN: the standard is appropriately not applied), false positives (FP: the standard is inappropriately applied) and false negatives (FN: the standard is inappropriately not applied). Measuring these yields measures of sensitivity and specificity. Measuring these at different values of a suitable technomarker yields a ROC curve for the application of a standard.*

To obtain a more holistic estimate of task-standard fit, SDT utilizes the area under the ROC curve (AUC). The ROC curve is generated by measuring precision and recall at various cut-off values. Selecting the optimum point on the ROC curve can calibrate standard application by identifying which technomarker cut-off value that provides the best discrimination between tasks suitable for standardised treatment and those that are not.

Comparing these measures for different variations of a standard, or different contexts or conditions of use, can guide research in optimisation of standard implementations, and guide local implementations seeking to calibrate the way they use standards. Although this description has focussed on the application of SDT to standards, the application should generalise to evaluating and calibrating the fit of any technology to a context.

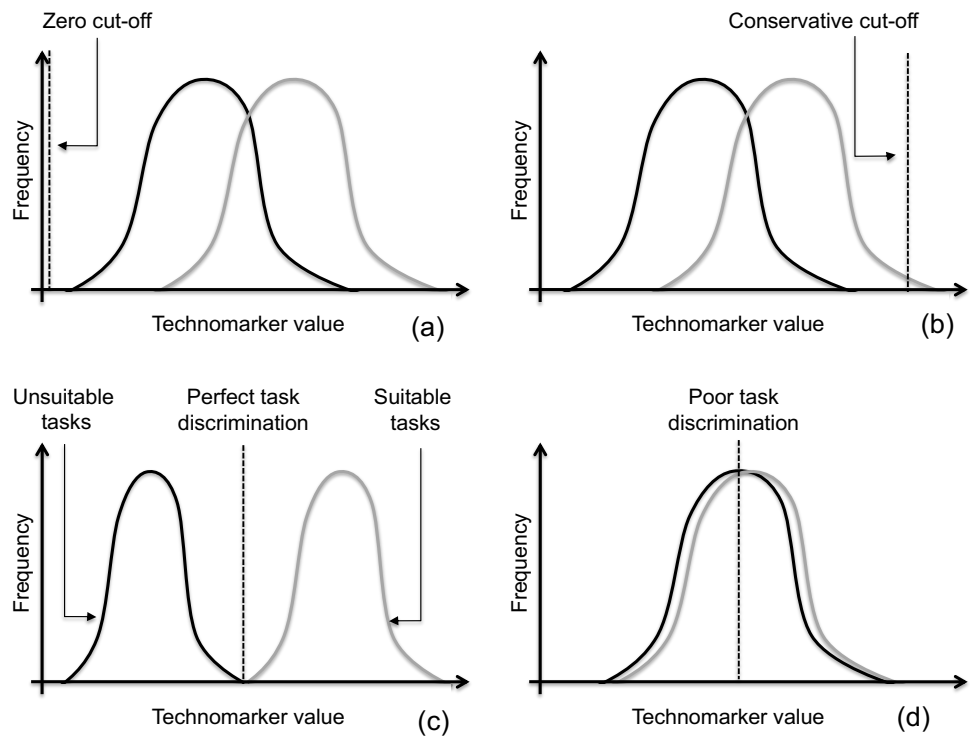

*Figure 3*: *The choice of cut-off value on technomarker affects the fitness of a standard to its context: (a) Variation agnostic: Ignoring the variable (setting the cut-off to zero) sees the standard indiscriminately applied across all tasks, triggering errors when there is variability in tasks.; (b) Variation avoidance: Choosing too conservative a value minimizes error at the cost of maximising the benefit of applying the standard. For a population of potentially distinguishable tasks, the choice of technomarker affects which tasks are considered suitable for standardisation: (c) A perfectly discriminating marker fully separates all tasks suitable for completion in a standardized way for those that are not; (d) A poorly discriminating marker choice makes it difficult to identify which tasks should be managed in a standardized way.*

*Measuring the impact of Standards on Outcomes*

Variability in design or implementation of a standard may lead to variations in the *outcome* of executing standardized actions. A standard's effectiveness can be estimated from the distribution of outcomes obtained when using it. The choice of outcome measure will depend on a standard's purpose e.g. a standardized process to diagnose patients could be measured by diagnostic accuracy, efficiency, or cost.

Comparisons can be made between an implemented standard's outcome distribution and a benchmark (such as a reference implementation). Such comparisons can yield fitness for purpose estimates including (Figure 4):
- *Intrinsic Fitness* (IF): The fitness that is associated with the shared common core between the compared distributions.
- *Additional Fitness* (AF): Any improvement in performance compared to the benchmark, for example through local improvements to a standard.
- *Missing Fitness* (MF): Any reduction in performance or fitness gap compared to the benchmark, for example because of local variations dependencies or resource constraints.

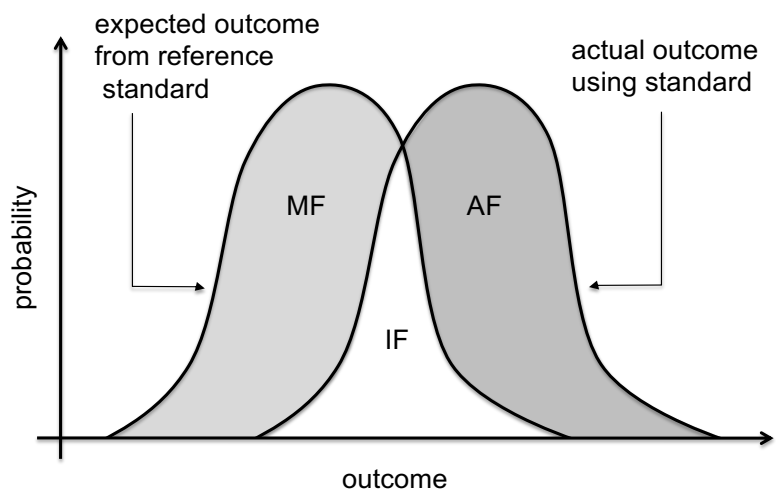

*Figure 4*: *The difference between expected and actual performance of an implementation of a standardized technology or process may be due to the additional fitness (AF) conferred (e.g. by variations in the local implementation) improving outcomes beyond those expected, or missing fitness (MF) where the implementation diminishes the likelihood of a desired outcome. Intrinsic fitness (IF) is associated with the shared common core.*

We could compare:
- The fitness of an implemented standard against an expected fitness codified in a gold standard reference implementation, moving conformance testing from fidelity of implementation to fidelity in performance.
- The fitness of an implemented standard against its historical performance.
- The fitness of two implementations of the same standard in different contexts.
- The fitness of two implementations of two different standards. (e.g. to help choose between HL7 and openEHR stacks[71])

We can define additional fitness measures in terms of AF, MF, and IF:
- *Potential Fitness* (PF) of a standard is its unrealized potential, and is the sum of its reference performance as well as any unexpected benefits from local modifications that could potentially be included in the standard i.e. PF = IF + AF + MF.
- *Realized Fitness* (RF) is the net achieved performance, calculated from the difference of fitness increases and decreases i.e. RF variation = AF − MF.

- *Fidelity of performance* (FI) measures the fit between a reference and implemented standard i.e. FI = IF – (AF + MF).

Fitness measures should be directly measurable from the areas under the standard performance curves. We can also choose to define a standard's fitness (F) as an expected utility, which is the product of the utility or outcome value (v) we assign to successfully completing a standardized task (ST), and the probability (p) of that event occurring i.e. F(ST) = v(ST) x p(ST).

**Disclosure Statements**:

Funding Statement: This work was supported by the National Health and Medical Research (NHMRC) Centre for Research Excellence in Digital Health, and NHMRC Investigator Grant GNT2008645.

Competing Interests Statement: EC is a co-founder, shareholder and Board member of Evidentli, a company that provides services and technologies to map information standards.

Contributorship Statement: EC conceived of and executed this study and wrote the paper.

Data Availability statement: Not applicable.